\documentstyle{mn}
\input{psfig}

\newif\ifAMStwofonts
\newcommand{\tP}{\tilde{P}}
\newcommand{\tF}{\tilde{F}}

\newcommand{\tA}{\tilde{A}}
\newcommand{\tB}{\tilde{B}}
\newcommand{\tS}{\tilde{S}}

\newcommand{\tx}{\tilde{x}}

\renewcommand{\bar}{\overline }

\newcommand{\xiav}{\bar{\xi}}
\newcommand{\cb}{\flat}

\ifoldfss
  \ifCUPmtlplainloaded \else
    \NewTextAlphabet{textbfit} {cmbxti10} {}
    \NewTextAlphabet{textbfss} {cmssbx10} {}
    \NewMathAlphabet{mathbfit} {cmbxti10} {} % for math mode
    \NewMathAlphabet{mathbfss} {cmssbx10} {} %  "   "    "
  \fi
  \ifAMStwofonts
    \ifCUPmtlplainloaded \else
      \NewSymbolFont{upmath} {eurm10}
      \NewSymbolFont{AMSa} {msam10}
      \NewMathSymbol{\upi}     {0}{upmath}{19}
      \NewMathSymbol{\umu}     {0}{upmath}{16}
      \NewMathSymbol{\upartial}{0}{upmath}{40}
      \NewMathSymbol{\leqslant}{3}{AMSa}{36}
      \NewMathSymbol{\geqslant}{3}{AMSa}{3E}

      \let\leq=\leqslant 
      \let\geq=\geqslant \let\ge=\geqslant
    \fi
  \fi
\fi % End of OFSS

\ifnfssone
  \newmathalphabet{\mathit}
  \addtoversion{normal}{\mathit}{cmr}{m}{it}
  \addtoversion{bold}{\mathit}{cmr}{bx}{it}
  \newmathalphabet{\mathbfit} % math mode version of \textbfit{..}
  \addtoversion{normal}{\mathbfit}{cmr}{bx}{it}
  \addtoversion{bold}{\mathbfit}{cmr}{bx}{it}
  \newmathalphabet{\mathbfss} % math mode version of \textbfss{..}
  \addtoversion{normal}{\mathbfss}{cmss}{bx}{n}
  \addtoversion{bold}{\mathbfss}{cmss}{bx}{n}
  \ifAMStwofonts
    \ifCUPmtlplainloaded \else
      \UseAMStwoboldmath
      \makeatletter
      \new@mathgroup\upmath@group
      \define@mathgroup\mv@normal\upmath@group{eur}{m}{n}
      \define@mathgroup\mv@bold\upmath@group{eur}{b}{n}
      \edef\UPM{\hexnumber\upmath@group}
      \new@mathgroup\amsa@group
      \define@mathgroup\mv@normal\amsa@group{msa}{m}{n}
      \define@mathgroup\mv@bold\amsa@group{msa}{m}{n}
      \edef\AMSa{\hexnumber\amsa@group}
      \makeatother
      \mathchardef\upi="0\UPM19
      \mathchardef\umu="0\UPM16
      \mathchardef\upartial="0\UPM40
      \mathchardef\leqslant="3\AMSa36
      \mathchardef\geqslant="3\AMSa3E

      \let\leq=\leqslant 
      \let\geq=\geqslant \let\ge=\geqslant
    \fi
  \fi
\fi % End of NFSS release 1

\ifnfsstwo
  \DeclareMathAlphabet{\mathbfit}{OT1}{cmr}{bx}{it}
  \SetMathAlphabet\mathbfit{bold}{OT1}{cmr}{bx}{it}
  \DeclareMathAlphabet{\mathbfss}{OT1}{cmss}{bx}{n}
  \SetMathAlphabet\mathbfss{bold}{OT1}{cmss}{bx}{n}
  \ifAMStwofonts
    \ifCUPmtlplainloaded \else
      \DeclareSymbolFont{UPM}{U}{eur}{m}{n}
      \SetSymbolFont{UPM}{bold}{U}{eur}{b}{n}
      \DeclareSymbolFont{AMSa}{U}{msa}{m}{n}
      \DeclareMathSymbol{\upi}{0}{UPM}{"19}
      \DeclareMathSymbol{\umu}{0}{UPM}{"16}
      \DeclareMathSymbol{\upartial}{0}{UPM}{"40}
      \DeclareMathSymbol{\leqslant}{3}{AMSa}{"36}
      \DeclareMathSymbol{\geqslant}{3}{AMSa}{"3E}

      \let\leq=\leqslant 
      \let\geq=\geqslant \let\ge=\geqslant
    \fi
  \fi
\fi % End of NFSS release 2

\ifCUPmtlplainloaded \else
  \ifAMStwofonts \else % If no AMS fonts
    \def\upi{\pi}
    \def\umu{\mu}
    \def\upartial{\partial}
  \fi
\fi

\title{Experimental Cosmic Statistics II~: Distribution }
\author[I. Szapudi et al.]{Istv\'an Szapudi,$^{1,2}$
St\'ephane Colombi,$^{3}$ 
Adrian Jenkins$^1$ \& J\"org Colberg$^4$ \\
$^1$University of Durham, Department of Physics, South Road,
Durham, DH1 3LE, UK\\
$^2$Canadian Institute of Theoretical Astrophysics,
60 St George St, Toronto, Ontario, M5S 3H8 Canada (present address)\\
$^3$Institut d'Astrophysique de Paris, CNRS, 98bis bd Arago,
F-75014 Paris, France\\
$^4$Max-Planck-Institut f\"ur Astrophysik, D-85740, Garching, Germany
}

\date{Submitted to MNRAS}
\pagerange{\pageref{firstpage}--\pageref{lastpage}}
\pubyear{1998}
\begin{document}
\voffset -0.5cm
\maketitle
\label{firstpage}
\begin{abstract}
Colombi et al. 1999 (paper I) 
investigated the counts-in-cells statistics
and their respective errors in the $\tau$CDM Virgo Hubble Volume
simulation. This extremely large $N$-body experiment also allows
a numerical investigation of the {\em cosmic distribution function}, 
$\Upsilon(\tA)$ itself for the first time.
For a statistic $A$, $\Upsilon(\tA)$ is the probability density of
measuring the value $\tA$ in a finite galaxy catalog.  $\Upsilon$ was
evaluated for the distribution of counts-in-cells, $P_N$, the factorial
moments, $F_k$, and the cumulants, $\xiav$ and $S_N$'s, using the same
subsamples as paper I.

While paper I concentrated on the first two moments of
$\Upsilon$, i.e. the mean, the cosmic error and the cross-correlations, here
the function $\Upsilon$ is studied in its full generality, including
a preliminary analysis of joint distributions $\Upsilon(\tA,\tB)$.
The most significant, and reassuring result for the
analyses of future galaxy data is that
the cosmic distribution function is nearly Gaussian provided 
its variance is small. A good practical criterion for
the relative cosmic error is that $\Delta A/A \la 0.2$. 
This means that for accurate measurements, 
the theory of the cosmic errors, presented by Szapudi \& Colombi (1996)
and Szapudi, Colombi \& Bernardeau (1999), and confirmed
empirically by paper I, is sufficient for a full statistical
description and thus for a maximum likelihood rating of models.
As the cosmic error increases, the cosmic distribution
function $\Upsilon$ becomes increasingly skewed and is well described
by a generalization of the lognormal distribution. The cosmic skewness
is introduced as an additional free parameter.
The deviation from Gaussianity 
of $\Upsilon(\tF_k)$ and $\Upsilon(\tS_N)$ 
increases with order $k$, $N$, and  similarly
for $\Upsilon(\tP_N)$ when $N$ is far from the maximum of 
$P_N$, or when the scale approaches the size of 
the catalog.  For our particular experiment, 
$\Upsilon(\tF_k)$ and $\Upsilon(\tilde{\bar \xi})$
are well approximated with the standard lognormal distribution,
as evidenced by both the distribution itself, and the comparison
of the measured skewness with that of the lognormal distribution.
\end{abstract}
%
%
%=======================
\section{Introduction}
%=======================
%
%
Precision higher order statistics will
become a reality when the new wide field surveys,
such as the SDSS and the 2dF, become available in the near future.
These prospective measurements contain information relating  to the 
regime of structure formation, to the nature of
initial conditions, and to the physics of galaxy formation.
The ability of such measurements to  constrain models, in a broad sense, 
is inversely proportional to the
overlap between the distribution of statistics predicted
by different theories for a finite galaxy survey. More precisely,
maximum likelihood methods give the probability of the particular measurements
for each theory, or after inversion, the likelihood of the
theories themselves. This is an especially natural and fruitful procedure
for a Gaussian distribution, where the first
two moments are sufficient for a full statistical description. 
This simple case is assumed for most analyses in the literature, 
and it motivates the special attention given to 
the investigation of the errors, or standard deviations.
In general, however, the underlying distribution of measurements can be strongly
non-Gaussian, in which case the correct shape for the distribution has
to be employed for a maximum likelihood analysis. As a consequence,
terms such as ``$1$-$\sigma$ measurement''
loose their usual meaning:  a few $\sigma$ deviation from the average can
be quite likely for a non-Gaussian distribution with a long tail.
Therefore it is of utmost importance to ask two important questions: 
\begin{enumerate}
\item[(i)] In what regime is the Gaussian approximation valid for the distribution
of the measured statistical quantities? 
\item[(ii)] If the Gaussian limit
is violated, is there any reasonably simple, practical assumption
which would enable a maximum likelihood
analysis? 
\end{enumerate}
This paper attempts to answer these questions by
studying numerically the underlying distribution function of measurements
for estimators of higher order statistics based on counts-in-cells. 
This complements the thorough
numerical investigation of the errors undertaken by Colombi et al.~(1999,
hereafter paper I), and the theoretical investigation of the
errors exposed in a suite of papers by Szapudi \& Colombi 
(1996, hereafter SC), Colombi, Szapudi, \& Szalay (1998, hereafter CSS),
and Szapudi, Colombi, \& Bernardeau (1999, hereafter SCB).

For a particular statistic $A$, $\Upsilon(\tA)$ denotes
the probability density of measuring a value $\tA$ in
a finite galaxy catalog. We consider the following counts-in-cells
statistics: factorial moments $F_k$, cumulants $\xiav$ and
$S_N$, void probability $P_0$ and its corresponding scaling
function $\sigma\equiv -\ln(P_0)/F_1$, as well the counts-in-cells distribution 
itself, $P_N$. A large $\tau$CDM 
$N$-body experiment, ${\cal E}$, generated by the VIRGO consortium (e.g., 
Evrard et al.~1999) was divided into $C_{\cal E}=4096$ 
cubic subsamples, ${\cal E}_i$, $i=1,\ldots,C_{\cal E}$ for estimating
numerically the cosmic distribution function, $\Upsilon(\tA)$. This was rendered
possible by the fact that this  ``Hubble Volume''
simulation involves $10^9$ particles in a cubic box of size $2000h^{-1}$ Mpc.
A detailed description of the simulation and the method we
used to extract count-in-cells statistics in the full box ${\cal E}$ and
its each of subsamples ${\cal E}_i$ can be found in paper I. 

Paper I concentrated entirely on the first two moments of 
$\Upsilon(\tA)$, the average
\begin{equation} 
 \langle \tA \rangle = \int \tA \Upsilon(\tA)d\tA ,
  \label{eq:mean}
\end{equation}
and the cosmic error
\begin{equation}
  (\Delta A)^2 \equiv \langle (\tA-\langle \tA \rangle)^2 \rangle = 
  \int (\tA - \langle \tA \rangle)^2\Upsilon(\tA)d\tA .
  \label{eq:variance}
\end{equation}
In the equations above, 
the mean $\langle \tA \rangle$ can
differ from the true value. The cosmic bias is defined as
\begin{equation}
 b_A \equiv \frac{\langle \tA \rangle}{A}-1.
\end{equation} 
It is always present when indicators are constructed
from unbiased estimators in a nonlinear fashion,
such as cumulants (e.g., SBC; Hui \& Gazta\~naga 1998, hereafter HG). 

The most relevant results of paper I are summarized next:
\begin{enumerate}
\item[(i) {\it The measured average $\langle \tA \rangle$}]
is in excellent agreement with perturbation theory, 
one-loop perturbation theory and extended perturbation 
theory (EPT) in their respective range of applicability. These
tests demonstrate the quality of our numerical experiment.

\item[(ii) {\it The measured cosmic error $\Delta A/A$}] is
in accord with the theoretical predictions 
of SC and SBC in their respective domain of validity.
A few percent accuracy is achieved in the weakly non-linear regime 
for the factorial moments. 
On small scales
the theory tends to overestimate the errors, perhaps
by a factor of two in the worst case, due to the
approximate nature of the  hierarchical models 
representing the joint moments (SCB).

\item[(iii) {\it The cosmic bias}] is
negligible compared to the errors in the full dynamic range,
as predicted by theory (SCB,
see also HG for an opposing view).

\item[(iv) {\it Cross-correlations between statistics of order
$k$ and $l$}] are in general agreement with
theory considering the preliminary nature of the measurements.
The precision of the predictions, however, decreases
with increasing difference of orders, $|k-l|$.
This suggests that the local Poisson
model (SC) looses accuracy, as expected.
\end{enumerate}

The theory of the errors confirmed by paper I
provides an excellent basis for future maximum likelihood
analyses of data whenever $\Upsilon$ is Gaussian.
While this was tacitly assumed by 
most previous works,
this article examines for the first time the range of validity of
this assumption. To this end the cosmic distribution 
function $\Upsilon(\tA)$ is examined numerically. In particular,
one of the parameters determining
its shape, the cosmic skewness
\begin{equation}
  S\equiv \langle (\tA -\langle \tA \rangle )^3 \rangle/(\Delta A)^3,
 \label{eq:cskew}
\end{equation}
is calculated as well.
When Gaussianity is no longer a good
approximation, new Ans\"atze are proposed for characterizing
$\Upsilon(\tA)$. 
In addition we perform a preliminary analysis of the bivariate cosmic distributions
$\Upsilon(\tA,\tB)$.

%Our results are organized as follows. 
The next section presents the estimates of $\Upsilon$ 
for the factorial moments, the cumulants (including the variance
of the counts),
the void probability distribution and its scaling function,
and the counts-in-cells themselves. A universal shape is found 
for $\Upsilon(\tA)$ which is well described in all
regimes by a generalized version of the lognormal distribution.
In addition to the mean (\ref{eq:mean}) and variance 
(\ref{eq:variance}), this depends on a third
parameter, the cosmic skewness (\ref{eq:cskew}). 
This is also investigated along
with the resulting {\em effective cosmic bias}. Section
3 presents the measured bivariate
distributions, with explicit comparison to theoretical predictions
of SCB. Finally, section 4
discusses the results in the context of maximum likelihood
analysis of future surveys. Readers unfamiliar
with counts-in-cells statistics 
can consult Appendix~A in paper I for a concise summary
of definitions and notation.
%
%
%========================
\section{The Cosmic Distribution Function}
%========================
%
%
The main results of this section are displayed
in figures 1--6. For simplicity figures 1, 3, and 5
will be referred to as type D, displaying distributions, while
figures 2, 4, and 6 as type S, showing skewness. A general
description of each type is followed by results obtained
for the cosmic distribution of the factorial moments
(\S~2.1), cumulants (\S~3.2),  counts-in-cells (\S~2.3), and
void probability with its scaling function  
$\sigma$ (\S~2.4). The cosmic skewness and the resulting
effective bias are discussed in \S~2.5.

In all figures of type D,  the results are displayed in
a convenient system of coordinates. For any statistic 
$\tA$ the normalized 
quantity
\begin{equation}
  \tx_A \equiv \frac{\delta \tA}{\Delta A} = 
  \frac{\tA -  A }{\Delta A}
  \label{eq:txa}
\end{equation}
is considered where $A=\langle \tA \rangle$ to simplify notations. 
The average of $\tx_A$ is 
zero and its variance is unity by definition which facilitates
the comparison of the plots. The disadvantage of this coordinate system is that
the cosmic error $\Delta A/A$ is not directly shown.

For reference, each figure of type D
displays a Gaussian (solid curve),
and  lognormal distribution with the same variance and average 
(dots, e.g.~Coles \& Jones 1991):
\begin{equation}
  \Upsilon(\tA)=\frac{1}{\tA  \sqrt{ 2\pi \kappa } }
  \exp\left\{ - \frac{ [\ln(\tA/A)+\kappa/2]^2}{2 \kappa}
  \right\},
\label{eq:lognor}
\end{equation}
with
\begin{equation}
  \kappa = \ln[1+ (\Delta A/A)^2 ].
\end{equation}
The skewness  of this distribution is given by
\begin{equation}
   S=(\Delta A/A)^3 + 3  \Delta A/A.
   \label{eq:slognor}
\end{equation}
For comparison, the skewness of the lognormal assumption 
is plotted with dotted lines on figures of type S.
The amount of skewness of the lognormal is a function of
the cosmic error, i.e. more skewness on the figures indicates
a larger cosmic error which is hidden by the choice of the
coordinate system.

In addition, a ``generalized lognormal distribution'' is introduced (dashes on
figures of type D):
\begin{eqnarray}
  \Upsilon(\tA) & = & \frac{s}{\Delta A [s (\tA-A)/\Delta A+1]  \sqrt{ 2\pi \eta } } \nonumber \\
  & & \times \exp\left( - \frac{ \{ \ln[s (\tA-A)/\Delta A+1]+\eta/2 \}^2}{2 \eta}
  \right),
\label{eq:lognor2}
\end{eqnarray}
\begin{equation}
 \eta=\ln(1+s^2),
\end{equation}
where $s$ is an adjustable parameter. It is fixed by
the requirement that the analytical function
(\ref{eq:lognor2}) have identical average, 
variance, and skewness,
$S = s^3 + 3 s$, with the measured $\Upsilon(\tA)$. 
It has more parameters, thus form (\ref{eq:lognor2}) 
characterizes the shape of function $\Upsilon(\tA)$ better than the
other two functions, especially for the large $\delta \tA$ tail. 
As will be shown next, it is an excellent approximation
for the underlying probability distribution {\em in all regimes
for all statistics}. This robust universality is the most striking
result of this article.

The  cosmic distribution function, as with any measurement from finite data,
is subject to both measurement and cosmic errors 
(the ``error on the error problem'', cf. SC).
The measurement error on $\Upsilon$, due to the finite
number of subsamples extracted from the
whole simulation, can be calculated
via straightforward error propagation. It essentially
corresponds to the usual $1/\sqrt{C_{\cal E}}$ factor,
where $C_{\cal E}$ is the number of subsamples. This is plotted
on all figures of type D as errorbars. On figures of type S
no errorbars are shown, since this would require an accurate estimate
up to the 6th moment of the cosmic distribution $\Upsilon(\tA)$. 
The excellent agreement 
between cosmic error measurements and theory (paper I)
indicates that the number of subsamples is
sufficient and thus the resulting errorbars should be fairly small.
Similar arguments
suggest that the simulation volume was sufficient large to render
the cosmic error on the cosmic distribution negligible. 
%========================
\subsection{Factorial Moments}
\begin{figure*}%[tbp]
\centerline{\hbox{
\psfig{figure=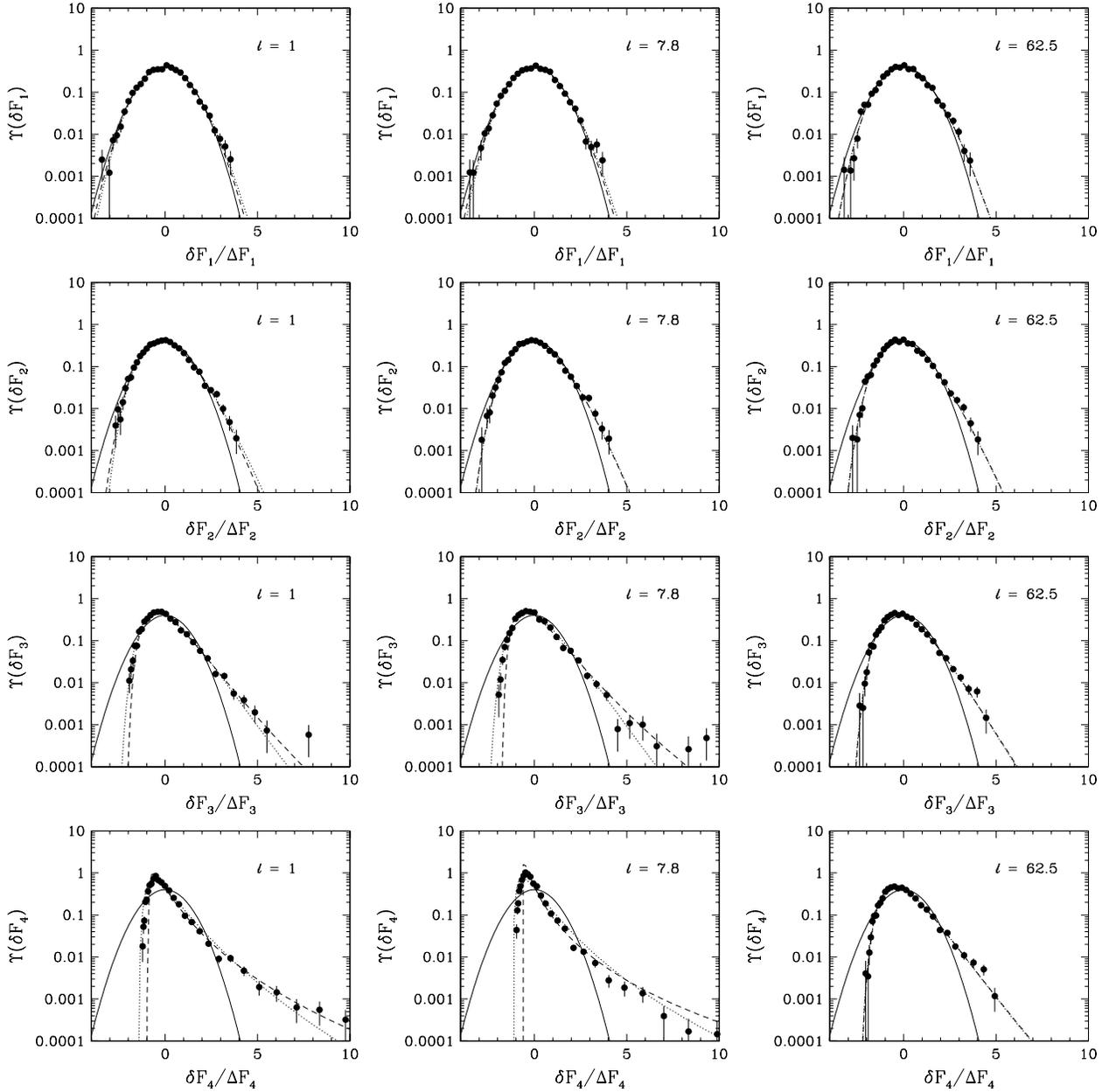,bbllx=45pt,bblly=186pt,bburx=574pt,bbury=708pt,width=17cm}}}
\caption[]{The cosmic distribution function of measurements 
$\Upsilon(\tF_k)$ shown as
a function of $\delta \tF_k/\Delta F_k$ 
as explained in the text. The scale of the measurements
$\ell=1$, $7.8$ and $62.5 h^{-1}$ Mpc is indicated on each panel. 
The order $k=1,2,3,4$  increases from top to bottom. 
The solid, dotted, and dash curves correspond to
the Gaussian, lognormal, and generalized lognormal
[eq.~(\ref{eq:lognor2})] distributions, 
respectively. While the coordinate system of the Figure 
does not display the value of the cosmic error directly,
the amount of skewness of the lognormal
distribution is an indicator of the magnitude 
$\Delta F_k/F_k$.
The errorbars show the measurement error
as discussed in beginning of \S~2.}
\label{fig:figure10}
\end{figure*}

%========================
Figure~\ref{fig:figure10} displays $\Upsilon(\tF_k)$ for $1 \leq k \leq 4$
and various scales $\ell =1, \ 7.8$, $62.5 h^{-1}$ Mpc. 

\begin{figure}%[tbp]
\centerline{\hbox{\psfig{figure=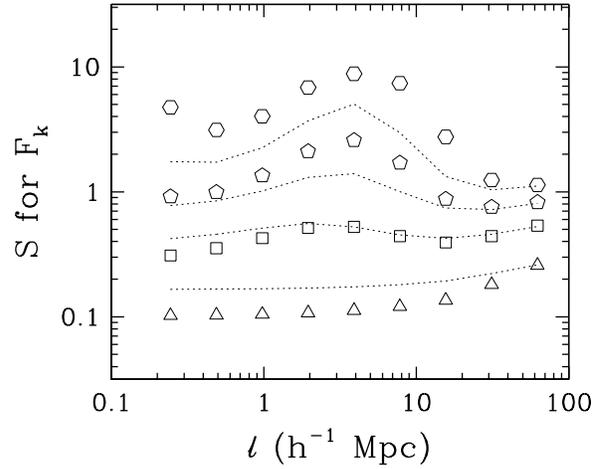,bbllx=72pt,bblly=120pt,bburx=536pt,bbury=490pt,width=8cm}}}
\caption[]{The skewness $S\equiv\langle (\tF_k-F_k)^3 
\rangle/(\Delta F_k)^3$ as a function of scale. The triangles, squares,
pentagons and hexagons respectively correspond to $k=1,2,3$ and $4$. There are
also dotted lines corresponding to an underlying
lognormal distribution (\ref{eq:slognor}); the orders increase
from bottom to top. The errors on the measurement have  not been
estimated since it would require a complicated calculation 
depending on the estimate of up to the $6^{\rm th}$ moment of $\Upsilon(\tF_k)$.
}
\label{fig:figure11}
\end{figure}

The agreement with the generalized  lognormal distribution
is excellent, but even the lognormal gives an adequate description.
The deviation from a Gaussian is pronounced whenever 
the relative  cosmic error $\Delta F_k/F_k$ is significantly
larger than unity. While the figures do not show the
cosmic error directly, the skewness of $\Upsilon(\tF_k)$ 
is a reliable indication. It increases with 
the order $k$ since $\Delta F_k/F_k$ also increases with $k$.
Figure~\ref{fig:figure11} shows directly the quantity $S$ measured for
$\Upsilon(F_k)$ along with the lognormal value (\ref{eq:slognor}).
The agreement shows that the lognormal model yields an excellent
approximation. 

Fig.~\ref{fig:figure10} in conjunction with
the measurements of the cosmic error in Paper I suggests that 
\begin{equation}
  \Delta A/A \la \Delta_{\rm crit}, \quad \Delta_{\rm crit}=0.2,
  \label{eq:rule}
\end{equation}
is a practical criterion for the validity of the Gaussian approximation. 

%========================
\subsection{Cumulants}
\begin{figure*}%[ptb]
\centerline{\hbox{
\psfig{figure=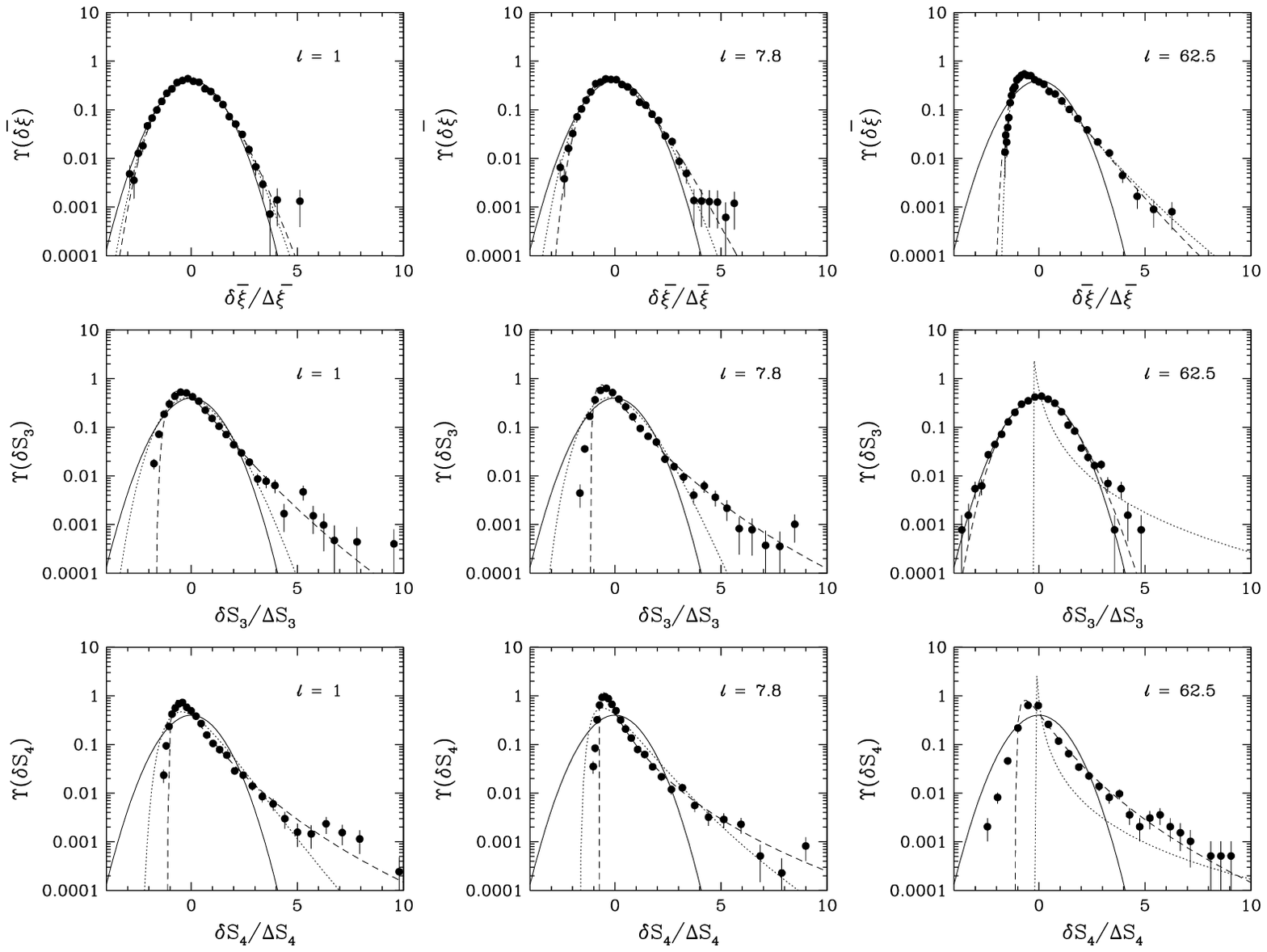,bbllx=84pt,bblly=327pt,bburx=539pt,bbury=675pt,width=17cm}}}
\caption[]{Analogous to Fig.~\ref{fig:figure10} for
$\Upsilon(\tilde{\xiav})$, $\Upsilon(\tS_3)$ and
$\Upsilon(\tS_4)$.}
\label{fig:figure12}
\end{figure*}

\begin{figure*}%[tbp]
\centerline{\hbox{\psfig{figure=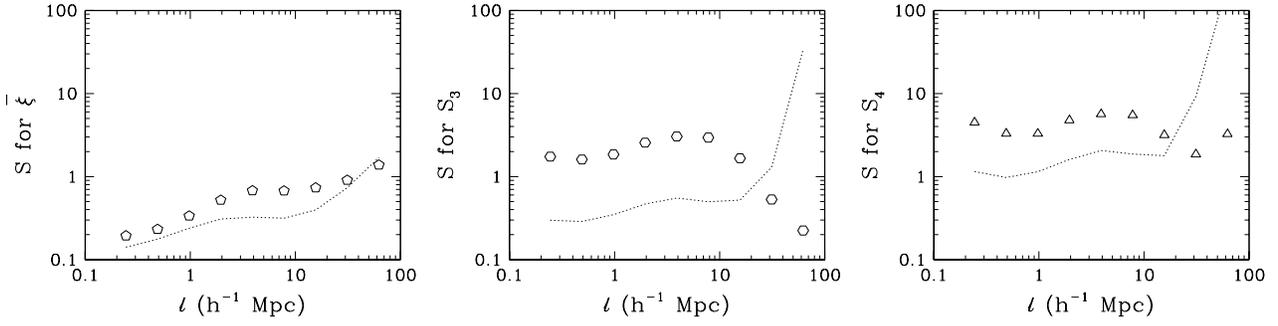,bbllx=78pt,bblly=551pt,bburx=538pt,bbury=675pt,width=17cm}}}
\caption[]{Same is in Fig.~\ref{fig:figure11}, but we consider here the skewness of $\tilde{\xiav}$ (left panel),
$\tS_3$ (middle panel) and $\tS_4$ (right panel).}
\label{fig:figure13}
\end{figure*}

%========================
Figure~\ref{fig:figure12} is analogous to Fig.~\ref{fig:figure10},
showing functions $\Upsilon(\tilde{\xiav})$, $\Upsilon(\tS_3)$ and
$\Upsilon(\tS_4)$ for the biased estimators. As was
shown in paper I, the bias is negligible compared to the cosmic
errors, thus correction is not necessary. 
The agreement with the lognormal is more approximate than
for $\Upsilon(\tF_k)$, except for the variance ${\xiav}$. 
Indeed, the skewness of $\Upsilon(\tS_N)$ is in general different
from the lognormal prediction, as illustrated by Fig.~\ref{fig:figure13}. 
On small scales it
is larger than predicted by equation~(\ref{eq:slognor}) 
while on large scales where edge effects
dominate it is much smaller. 
The generalized lognormal (\ref{eq:lognor2}) can still
account for the shape of $\Upsilon(\tS_N)$ quite well, 
especially for the large $\tS_N$ tail.

The cosmic skewness
of $\Upsilon(\tS_k)$ is fairly small on large scales. This is a natural
consequence of the fact that cumulants are not subject to the
positivity constraint $\tS_k \geq 0$,
as it is the case for factorial moments. 
On large scales, the
measured $\tS_k$ may well be positive or negative, similarly
with $\xiav$ on extremely large scales.
As a result,
the left-hand tail of  the distribution is more pronounced
in both lower right panels of Fig.~\ref{fig:figure12} than
the corresponding figure for factorial moments,  and  $\Upsilon(S_3)$
is almost Gaussian in the middle right panel. 

Rule (\ref{eq:rule})
for the Gaussian limit still applies, at least for $\xiav$, 
and perhaps  a slightly more stringent condition should be chosen 
for cumulants of higher order. $\Upsilon(\tS_3)$ is
fairly skewed even though the measured cosmic error is slightly 
below the threshold value for $\ell=1 h^{-1}$ Mpc and 
$\ell = 7 h^{-1}$ Mpc (see paper I).
%========================
\subsection{Counts-in-cells}
\begin{figure*}%[ptb]
\centerline{\hbox{
\psfig{figure=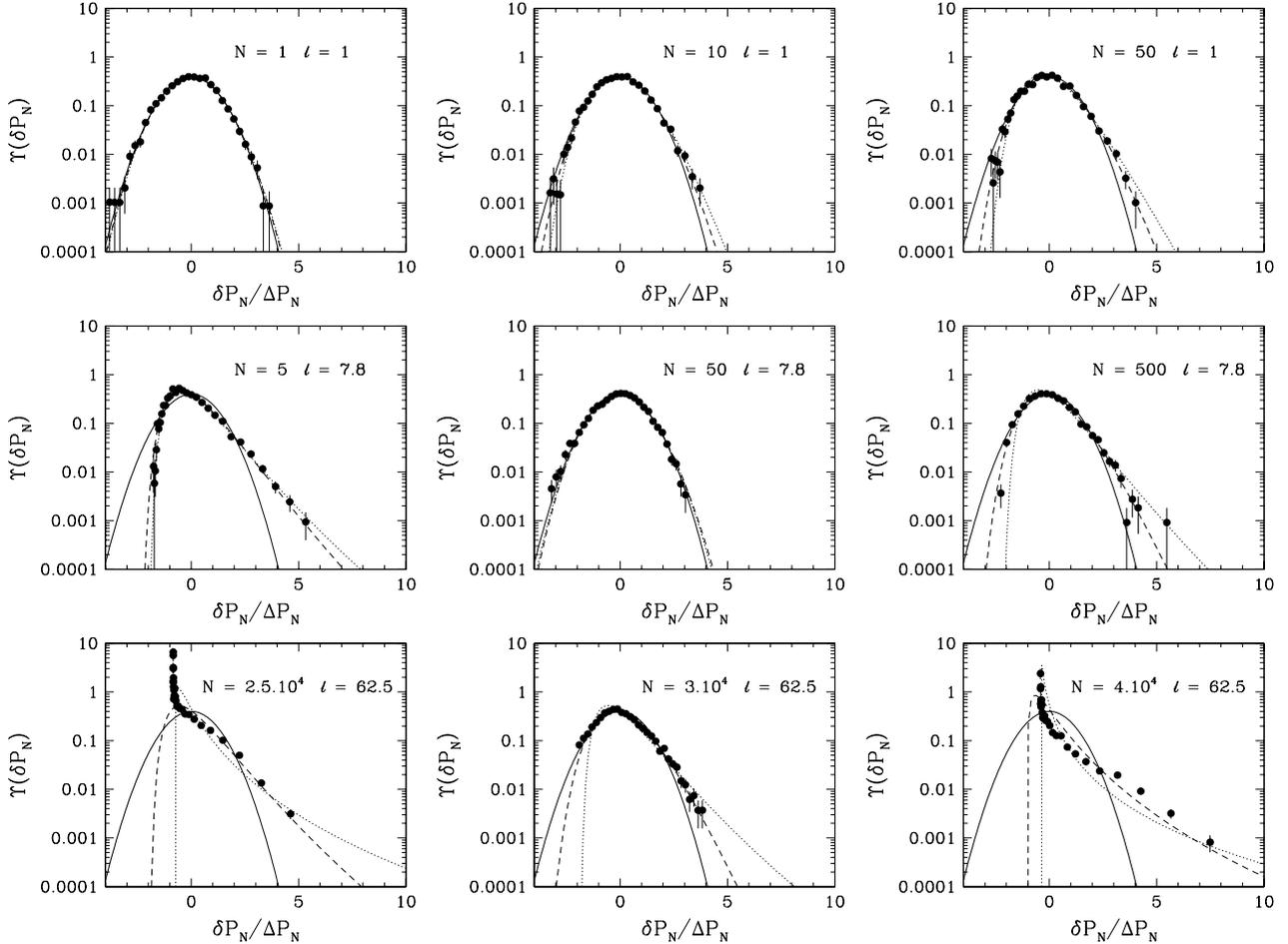,bbllx=54pt,bblly=317pt,bburx=574pt,bbury=706pt,width=17cm}}}
\caption[]{Same as in Fig.~\ref{fig:figure10}, but
now, the distribution function of measurements $\Upsilon(\tP_N)$ is shown as
a function of $\delta \tP_N/\Delta P_N$  for various scales and values of $N$
as indicated on each panel.}
\label{fig:figure14}
\end{figure*}

\begin{figure}%[tbp]
\centerline{\hbox{\psfig{figure=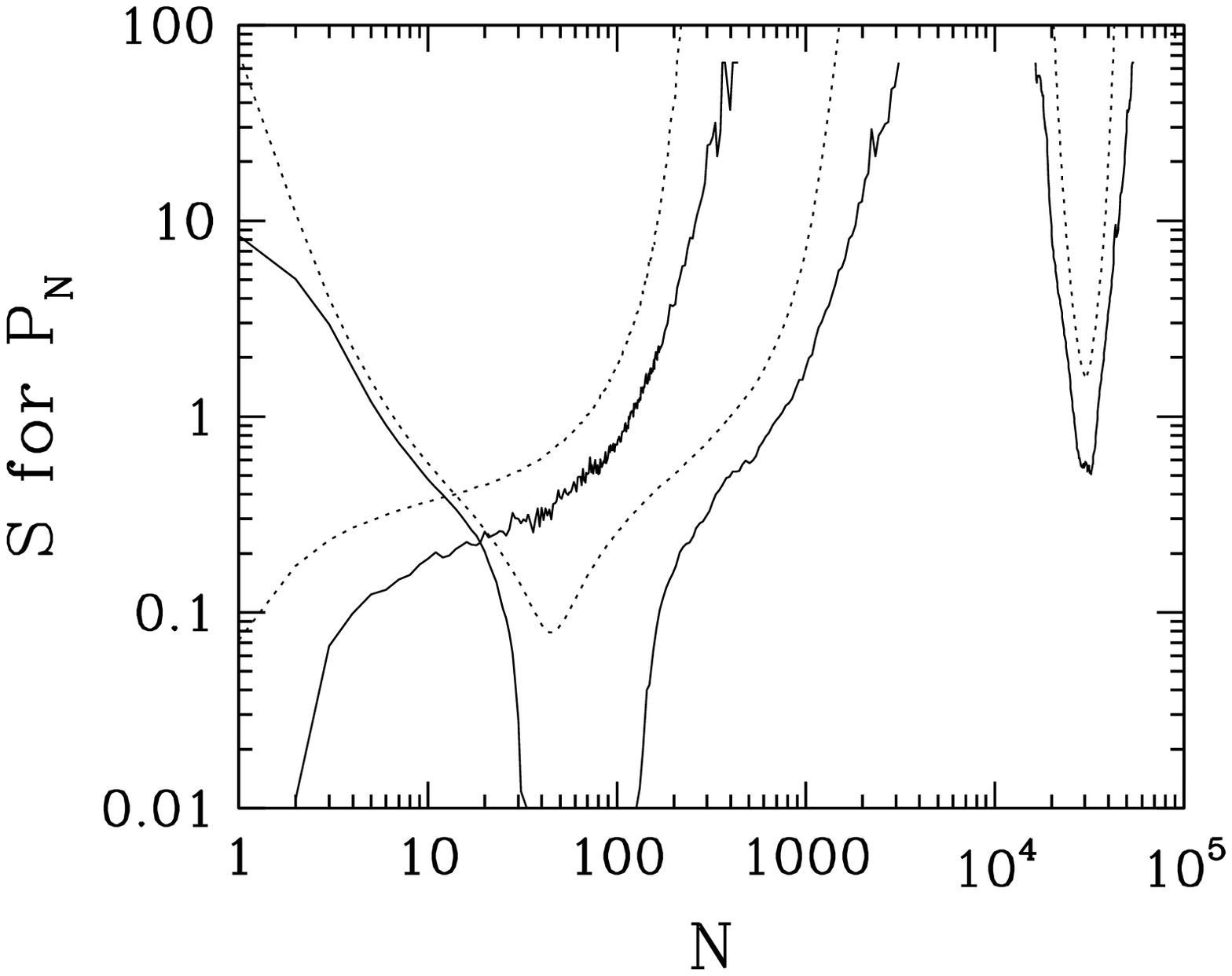,bbllx=66pt,bblly=120pt,bburx=529pt,bbury=493pt,width=8cm}}}
\caption[]{The skewness $S$ of $\Upsilon(\tP_N)$ as a function of $N$ for three scales, $\ell = 1 h^{-1}$ Mpc (left
curve), $\ell=7.8 h^{-1}$ Mpc (middle curve) and $\ell=62.5 h^{-1}$ Mpc (right curve). The dotted curves
give the lognormal prediction, which is always larger than the measurement. }
\label{fig:figure15}
\end{figure}

%========================
Figure~\ref{fig:figure14} shows the function $\Upsilon(\tP_N)$ in various
cases. 
The upper panels focus on a small scale $\ell \simeq 1 h^{-1}$ Mpc. 
In this regime, the CPDF and $-\Delta P_N/P_N$ are 
decreasing functions of $N$ as demonstrated in paper I.
%(third curve from the left
%in Fig.~\ref{fig:figure1} and in upper panel of Fig.~\ref{fig:figure8}).
Once again, the validity of the Gaussian approximation depends on the size of
cosmic error.
As a result, $\Upsilon(\tP_N)$ is nearly Gaussian 
for $N=1$ and becomes more and more
skewed as $N$ increases. 
The lognormal approximation appears to be adequate within the errors, 
although it is slightly too skewed as illustrated by Fig.~\ref{fig:figure15}.

The middle panels show an intermediate scale $\ell \simeq 7.8 h^{-1}$ Mpc.
On these scales (cf. paper I)
both the CPDF and the cosmic error have a unimodal behaviour with
an extremum (maximum for the CPDF and correspondingly minimum
for the errors) for $N \sim N_{\rm max} =26$. This explains why for the
chosen values  of $N=5$, $50$, and $500$,
function
$\Upsilon(\tP_N)$ is skewed, approximately Gaussian, and skewed again
respectively.
For $N=5$ lognormal is an excellent approximation, while
the skewness for $N = 500$ is somewhat less than that of a lognormal.

Finally, the lower panels display the largest available scale 
$\ell = 62.5 h^{-1}$ Mpc. The behaviour of $P_N$ and $\Delta P_N/P_N$ 
is similar as previously with the extremum shifted to
$N \sim N_{\rm max} \simeq 30000$. In this case, 
the cosmic error is always large,
at least of order fifty percent (cf. paper I).
All the curves are thus significantly skewed for the chosen
values of $N=25\,000$, $30\,000$ and $40\,000$. The agreement with
the lognormal assumption is somewhat inaccurate, 
although the generalized lognormal improves the fit, especially for
the left-hand panel. 
Note that the apparently abrupt limit for small values
of $\delta \tP_N/\Delta P_N$ is due to the positivity 
constraint $\tP_N \ge 0$. This constraint becomes
quite severe when the average value is much smaller
than the errors. While there is still plenty of dynamic range
for upscattering, there is a hard restriction for down scattering.
This is only partly taken into account in our generalized 
lognormal model, and any modifications in this respect
are left for future work.
Finally, the practical criterion (\ref{eq:rule}) is again valid
for determining Gaussian approximation.

%In contrast with Fig.~\ref{fig:figure10} where 
%$\tF_k=0$ corresponded to $\Upsilon(\tF_k=0)=0$, here
%$\Upsilon(\tP_N=0)$ can be finite or even infinite. 
%{\bf I have to think about this}This
%shows up as a sudden upturn of the curves and the corresponding
%points.

Note that the finite number $C=512^3$ of sampling cells (see paper I), the CPDF is
necessarily a multiple of $1/C$. This quantization could
cause contamination of 
$\Upsilon(\tP_N)$ unless $P_N \gg 1/C \simeq 10^{-8.13}$. 
The condition  $P_N \ge 10^{-6}$ adopted corresponds to at
least $\sim 100$ cells per subsample in average with $N$ particles.
Despite that, a small amount of contamination might still persist
for $\delta \tP_N \ga -P_N$,
i.e. at the left side of  the plots on figure~\ref{fig:figure14}. 
The same effect
might also alter the tail of the counts-in-cells measurements
presented in paper I, although not significantly.

%
%=============================================================
\subsection{Void Probability and Scaling Function}
%=============================================================
According to the investigations in paper I,
the cosmic error on $P_0$ and $\sigma$ 
increases steadily with scale up to a sudden transition  
on scales $\ell \sim 5 h^{-1}$ Mpc where it becomes large or infinite. 
This behavior was studied extensively by
CBS where more of the details can be found.
The most relevant consequence here is that
in the available dynamic range the cosmic error is small, 
and $\Upsilon(\tP_0)$ and $\Upsilon({\tilde \sigma})$ are nearly Gaussian.
For this reason it would be superfluous to print the corresponding
figures.

%
%

%
%
%=============================================================
\subsection{Cosmic Skewness and Cosmic Bias}
%=============================================================

According to Figs.~\ref{fig:figure10}--\ref{fig:figure15}, the 
degree of skewness of the cosmic distribution function 
increases with the order $k$ and with $|N-N_{\rm max}|$,
where $N_{\rm max}$ is the value for which $P_N$ reaches its maximum. 
The cosmic
skewness is already significant for third order statistics, $F_3$ and $S_3$. 
An important consequence of the large cosmic skewness is
that the maximum $\Upsilon(\tA)$, i.e. the most likely measurement,
is shifted to the left from the ensemble average on
Figs.~\ref{fig:figure10}, \ref{fig:figure12} and \ref{fig:figure14}. 
Maximizing the Ansatz (\ref{eq:lognor2}), 
which is always a good fit to the cosmic distribution function, 
yields 
\begin{equation}
 \cb_A=A_{\rm max}/A-1=\frac{\Delta A}{A s} \left( \frac{1}{(1+s^2)^{3/2}}- 1 \right),
\end{equation}
where $\cb_A$ is the {\em effective } cosmic bias. 
Since $s > 0$, it is negative, and its absolute value is
smaller than the cosmic error, 
\begin{equation}
	 |\cb_A| \la 0.66\frac{\Delta A}{A}.
\end{equation}
For a lognormal distribution, $s=\Delta A/A$,
\begin{equation}
  |\cb_A|=1-\left[ 1 +(\Delta A/A)^2 \right]^{-3/2} \leq 1.
  \label{eq:biasln}
\end{equation}
The effective cosmic bias becomes increasingly significant
when the cosmic error is large.
Similarly to the cosmic bias (SBC), $\cb_A \sim -(3/2) (\Delta A/A)^2$ 
from expanding eq.~(\ref{eq:biasln}) in the small error regime.

The phenomenon of effective bias was already
pointed out by SC (and preliminarily investigated by
Colombi, Bouchet \& Schaeffer, 1994).
Since $A_{\rm max}$ is the most likely value
of $\tA$, the only one available measurement 
in a catalog of the neighbouring Universe
is likely to yield lower than average value.
This is true even
for an unbiased indicator such as $\tF_k$ or $\tP_N$. 
Unfortunately, this effect cannot
be corrected for, but it can be taken into account
in the framework of the maximum likelihood approach
using the above results on the shape of $\Upsilon(\tA)$.

%
%=============================================================
\section{Bivariate Cosmic Distribution Function: a Preliminary Analysis}
\begin{figure*}%[ptb]
\centerline{\hbox{
\psfig{figure=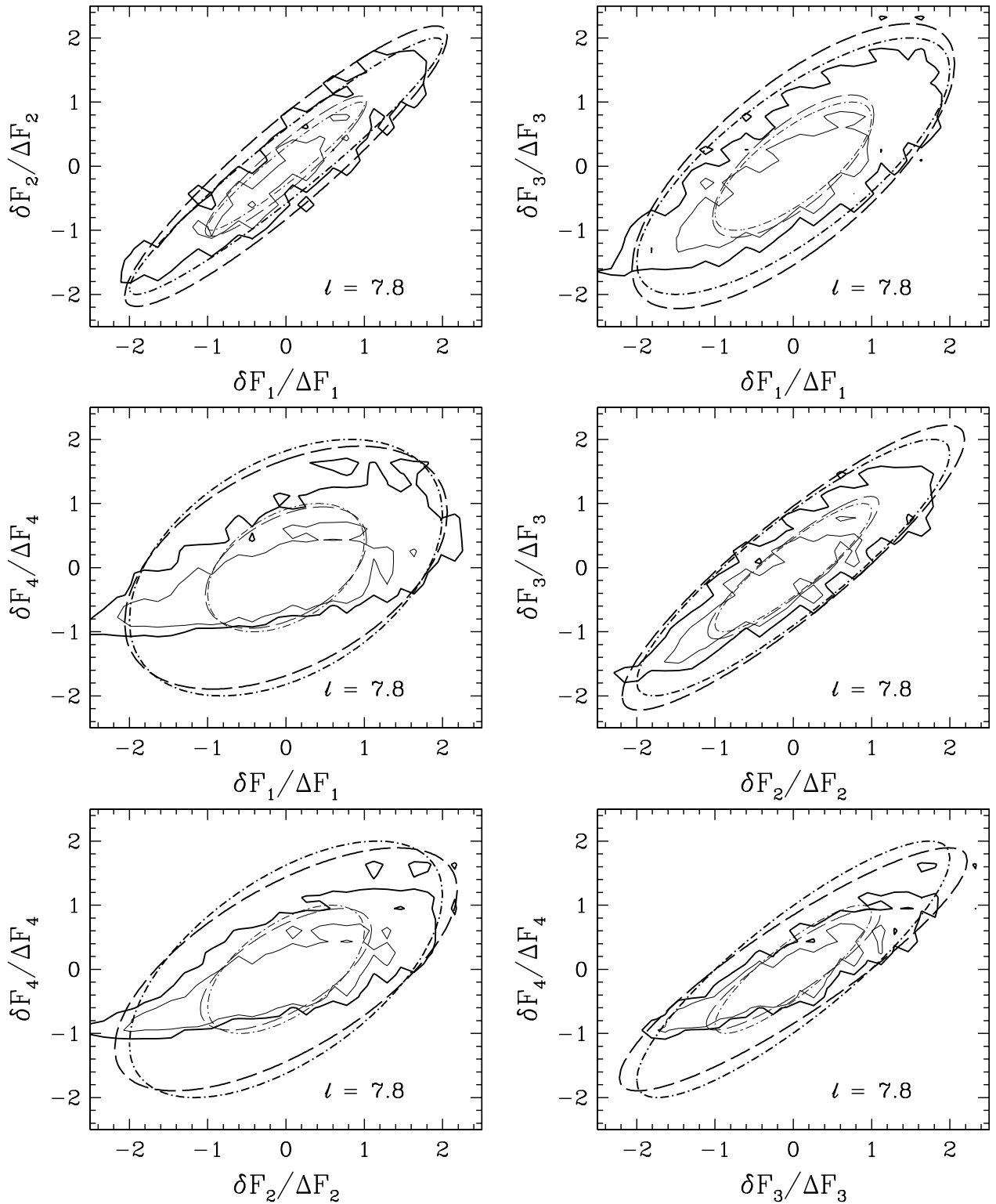,bbllx=60pt,bblly=85pt,bburx=573pt,bbury=706pt,width=17cm}}}
\caption[]{The joint cosmic distribution function for factorial moments,
$\Upsilon(\tF_k,\tF_l)$. Thin and thick solid contours 
are displayed for two values of
$\Upsilon$  which would correspond respectively
to $1\sigma$ and $2\sigma$ contours in the Gaussian limit.
The latter is shown
as thin and thick dot-dashes. For comparison, the analytic prediction
of SCB for E$^2$PT is also plotted with thin and thick long dashes corresponding
to the
Gaussian limit with theoretical cosmic errors and cross-correlation
coefficient. The scale of the measurement is $\ell=7.8\ h^{-1}$ Mpc 
as displayed on each panel. The image used to draw contour
plots has $30^2$ pixels. It was generated using bilinear interpolation from
an other array with logarithmic binning in each coordinate 
in order to reduce the errors on the estimate of function 
$\Upsilon(\tA,\tB)$ in each bin.}
\label{fig:figurebiv1}
\end{figure*}

\begin{figure*}%[ptb]
\centerline{\hbox{
\psfig{figure=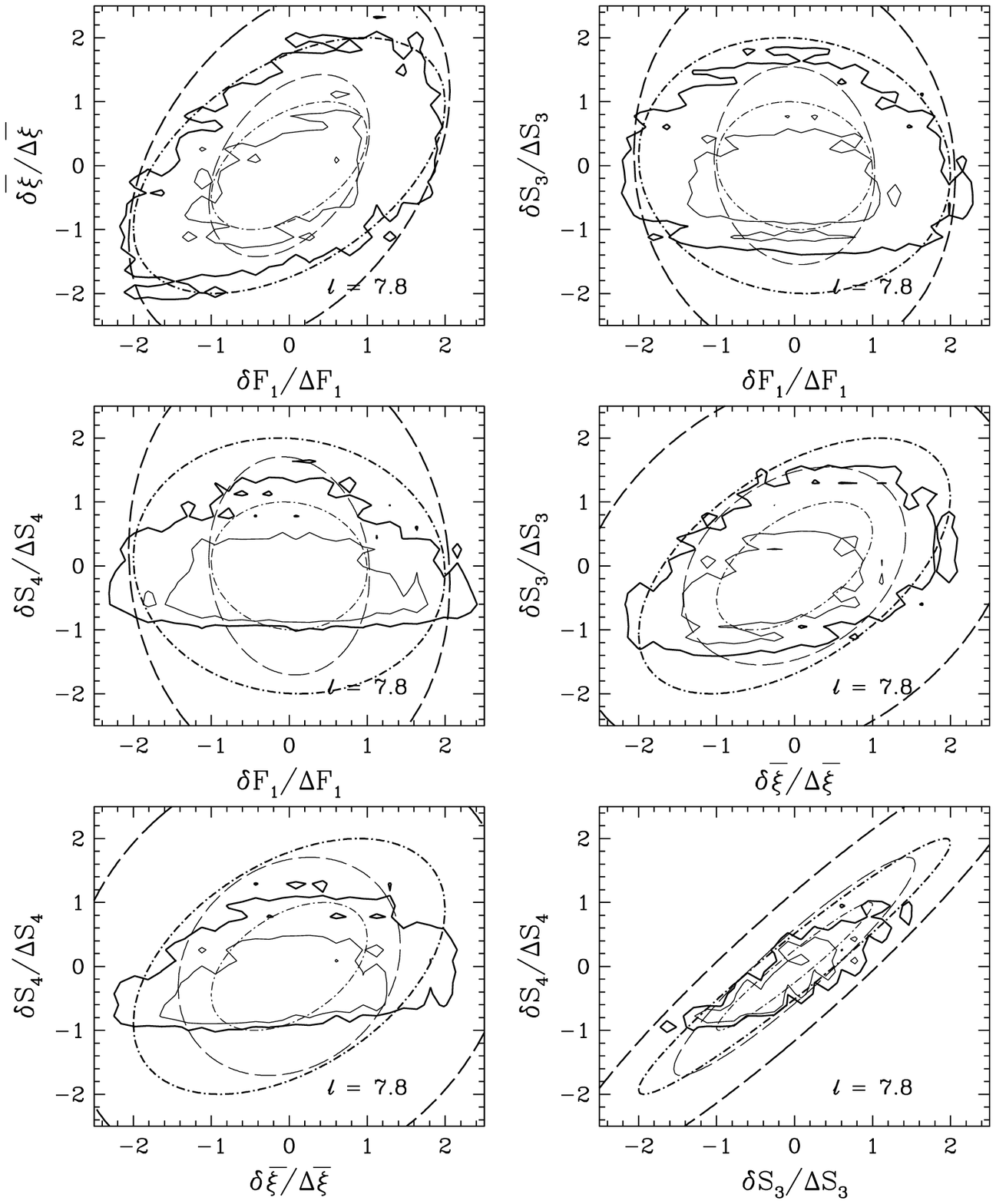,bbllx=91pt,bblly=128pt,bburx=537pt,bbury=671pt,width=17cm}}}
\caption[]{Same as in Fig.~\ref{fig:figurebiv1}, but for the average count $F_1$ and
the cumulants, $\xiav$, $S_3$ and $S_4$.}
\label{fig:figurebiv2}
\end{figure*}

%=============================================================
%
Figures~\ref{fig:figurebiv1} and \ref{fig:figurebiv2} display 
contours of the joint cosmic distribution $\Upsilon(\tA,\tB)$ 
(solid lines) for factorial moments and cumulants, respectively.
For comparison the Gaussian limit is shown,
\begin{equation}
   \Upsilon(\tA,\tB)=\frac{1}{2\pi \Delta A  \Delta B \sqrt{1-\rho^2}} \exp \left[-
  \frac{1}{2}{\cal Q}(\tA,\tB) \right],
\end{equation}
\begin{equation}
  {\cal Q}(\tA,\tB)=\frac{1}{1-\rho^2} \left[ 
 \tx_A^2 -2 \rho \tx_A \tx_B + \tx_B^2 \right],
\end{equation}
where $\rho\equiv \langle \delta \tx_A \delta \tx_B \rangle$
is the cross-correlation coefficient. Dot-dashes display
the above function with the measured
$\rho$, $\Delta A/A$ and $\Delta B/B$, while long dashes represent
the same function but with the parameters inferred from the theory of SCB 
with the E$^2$PT model (see paper I for details). 
The contours, correspond in the Gaussian limit
to the 1$\sigma$ (thin curves) level, 
${\cal Q}(\tA,\tB)=1$, and the 2$\sigma$ (thick curves) level, 
${\cal Q}(\tA,\tB)=4$, are displayed in the coordinate system of the measured
$\tx_A$ and $\tx_B$. 

On $\ell=7.1 h^{-1}$ Mpc scales
the theoretical predictions are expected
to match the second order moments of 
$\Upsilon$ for factorial moments, and
even the cross-correlations (see Paper I).  This is illustrated by 
Fig.~\ref{fig:figurebiv1}, where the long-dashed ellipses superpose well to
the dot-dashed ones. For the cumulants the theory overestimates
the errors slightly, which is reflected in the contours of
Fig.~\ref{fig:figurebiv2},
although cross-correlations are still reasonable, as indicated by
the orientation of the ellipses. 

The departure from the Gaussian limit is significant,
except for the upper left panel on Figs.~\ref{fig:figurebiv1}
and \ref{fig:figurebiv2}, and increases with order, in accord
with the findings of the previous section.
The contrast with Gaussianity  increases
with the cosmic error, 
and thus with the order considered. With the exception of ${\bar N}$,
$F_2$, $\xiav$ and $S_3$, the measured cosmic error violates
(\ref{eq:rule}) at $\ell=7.1 h^{-1}$ Mpc (see paper I). 
Moreover, as shown previously, criterion 
(\ref{eq:rule}) should be strengthened
for cumulants $S_k$, $k \geq 3$.
% (but note however that departure
%from Gaussianity is not very 
%pronounced in upper right and middle right panels 
%of Fig.~\ref{fig:figurebiv2}). 
In conclusion, condition (\ref{eq:rule}) distinguishes
the Gaussian limit for $\Upsilon(\tA,\tB)$ adequately 
when applied to both statistics $\tA$ and $\tB$.

Similarly to the monovariate distribution (\S~2),
function $\Upsilon(\tA,\tB)$ develops skewness and a
significant tail for large values of $\tx=(\tx_A,\tx_B)$
when rule (\ref{eq:rule}) is broken.
There are three notable consequences:
\begin{enumerate}
\item[(i)] The effective cosmic bias 
(\S~2.5) is present again, i.e. 
the maximum of $\Upsilon$ is shifted from the 
average towards the lower left corner of the panels.
\item[(ii)] The contours tend to cover a smaller area 
than for the Gaussian limit.
\item[(iii)] As a result of the positivity constraint, 
there is a  well defined lower vertical/horizontal bound
in some panels, e.g., for $\tx_{F_4}$,  $F_4\geq 0$.
\end{enumerate} 
%
%
%
%=============================================================
\section{Summary and Discussion}
%=============================================================
%
%
This paper has presented an experimental study of the cosmic distribution 
function of measurements $\Upsilon(\tA)$, where $\tA$ is an indicator of a
statistic related to counts-in-cells. The cosmic distribution was
considered for the factorial moments $F_k$, cumulants $\xiav$ and $S_N$, 
the void probability $P_0$ with its scaling function, 
$\sigma\equiv -\ln(P_0)/F_1$, and finally the counts-in-cells $P_N$ themselves. 
To analyse properties of the function $\Upsilon(\tA)$, we used 
a state of the art $\tau$CDM simulation divided into 4096 sub-cubes  
large enough themselves to represent a full galaxy catalog. 
The statistics mentioned above were extracted from each subsample, 
and the resulting distribution of measurements was used to 
estimate $\Upsilon(\tA)$.

While paper I concentrated on the first two moments of
the cosmic distribution, the average and the errors,
here the focus was shifted towards the general
shape of function $\Upsilon$ itself, including its skewness,
the cosmic skewness.
The main results of this analysis are the followings:
\begin{enumerate}
\item[(i)] In contrast with popular belief, the cosmic distribution
is {\em not Gaussian} in general. The most reassuring result
is, however, that the Gaussian approximation appears to 
be valid whenever the cosmic errors are small, typically
$\Delta A/A \la 0.2$. This result is quite
robust and it is insensitive to the particular statistic considered
(except that a slightly more stringent condition might be chosen
for cumulants $S_k$, $k \geq 3$).
This means that for any quantity which can be reliably measured
from a survey, a Gaussian error analysis should be valid.

When the relative cosmic error $\Delta A/A$ becomes significant, $\Upsilon$  
becomes increasingly skewed. Since
$\Delta F_k/F_k$ and $\Delta S_k/S_k$ increase with 
$k$ (SC, paper I), and $\Delta P_N/P_N$  with $|N-N_{\rm max}|$, where
$N_{\rm max}$ is the maximum of the CPDF, 
so does the cosmic skewness,
which eventually results in the break down of the Gaussian approximation.
Functions $\Upsilon(\tF_k)$  and $\Upsilon(\tilde{\xiav})$ are 
well approximated by a lognormal law. Otherwise,
a third order parametrisation matching the average, 
the variance and the
skewness of the observed distribution is necessary, 
and in general sufficient.
Such a generalization of lognormal distribution is proposed and
found to be in agreement with the measurements in all regimes
investigated. Note that there are other alternatives  
such as the Edgeworth expansion (e.g., Juszkiewicz et al.~1995) or
the skewed lognormal approximation of Colombi (1994).  This latter
consists of applying Edgeworth expansion to $\log(\tA)$. This method,
when applicable, improves significantly the domain of 
validity of the Edgeworth expansion, normally only useful in
the weakly non-Gaussian limit $\Delta A/A \la 0.5$. 

\item[(ii)] While paper I examined the cosmic bias resulting
from the non-linear construction of certain estimators, 
here a new phenomenon was pointed out, which is similar
in effect, but different in nature: the {\em effective cosmic
bias}. It affects all estimators, including unbiased ones,
and is a result of the cosmic skewness. Whenever the
cosmic errors are large, the cosmic distribution function
develops a skewness corresponding to a long tail. As a result,
the most likely measurement will be smaller than the average.
Such a phenomenon was pointed out earlier in SC, and here it
has been found to be universal. As SCB and paper I found that
the cosmic bias is usually insignificant compared to the
cosmic errors, it is likely that the effective cosmic
bias is responsible for some of the conspicuously low measurements
from small galaxy catalogs. This is in contrast with
the conjecture of Hui \& Gazta\~naga (1998, hereafter HG), 
who assumed that the cosmic bias resulting
from the use of biased estimators could explain this phenomenon.
The effective cosmic bias renders correction for the cosmic bias useless, in
contrast with the proposition of HG. The effective cosmic
bias (and the less significant cosmic bias if any) can be
taken into account in the framework of a full maximum likelihood
analysis, which relies on the shape of the cosmic
distribution function approximated with sufficient accuracy.

\item[(iii)] A preliminary investigation of joint distribution
$\Upsilon(\tA,\tB)$ was performed for factorial moments and cumulants. It
confirms the validity of the above points (i) and (ii) for cosmic
bivariate distribution.
In particular, a practical criterion 
for the validity of the Gaussian limit is that
the cosmic error for both estimators be small enough, typically
$\Delta A/A \la 0.2$ and $\Delta B/B \la 0.2$. This result 
can be safely generalized to $N$-variate distribution
functions, thus providing the basis of full multivariate
maximum likelihood analysis of data in the Gaussian limit.

We have not attempted to develop a more accurate multivariate
approximation than (multivariate) Gaussian as this would go beyond the
scope of this paper.  However, we conjecture that an extension of our
generalized lognormal distribution would be feasible (see the point of
view of Sheth, 1995).  An alternate approach, proposed by Amendola
(1996), would employ a multivariate Edgeworth expansion. However,
similarly with point (i) above for monovariate distributions, this
approximation is only valid when the errors are small; but this is
precisely the criterion for the Gaussian limit as we shown previously.  A
generalization of the lognormal distribution expanding the logarithm
of the statistics via the multivariate Edgeworth technique provides a
potential improvement of this method.
\end{enumerate}

It is worth noting that the behaviour of the cosmic distribution function
is expected to be extremely robust with respect to 
the particular model studied in this paper, $\tau$CDM.
For example, SC, in their preliminary investigations, found
essentially the same universal behaviour in Rayleigh-Levy fractals.
Moreover, as discussed more extensively in Paper I,
the results are sufficiently
stable that the usual worries of galaxy biasing
(not to be confused with cosmic and effective cosmic bias)
and redshift distortions are unlikely to change them
qualitatively. Indeed the shape of the cosmic distribution
function is almost entirely determined by the magnitude of the 
cosmic error, and it is insensitive to which  statistic
is considered. The powerful universality found among
entirely different statistics is likely to carry over
when the two effects mentioned above, which are subtleties in comparison 
with the range of statistics investigated, are taken into account.

The results found in the present work and in paper I
are encouraging for investigations 
in future large galaxy catalogs
and for problems related to data compression 
(e.g. Bond 1995; Vogeley \& Szalay 1996; 
Tegmark, Taylor \& Heavens 1996; Bond, Jaffe \& Knox 1998; Seljak 1998).
For example, the cosmic error on factorial moments
is expected to be small on a large dynamic range
in the SDSS (see, e.g.~CSS), implying according to the above findings that
the cosmic distribution function should be 
nearly Gaussian in this regime. In that case, 
theory of the cosmic errors and cross-correlations, outlined
in SC, CSS and SCB and thoroughly tested in paper I, 
will be sufficient for full multivariate maximum likelihood analyses.
Preliminary investigations on current surveys are
being undertaken by Szapudi, Colombi \& Bernardeau (1999b) 
and Bouchet, Colombi \& Szapudi (1999).
Similarly the  theoretical background is currently being developed
for future weak lensing surveys (Berneardeau, Colombi, Szapudi, 1999), 
where statistical analyses will be conducted with indicators
very close to counts-in-cells
(see, e.g. Bernardeau, Van Waerbeke \& Mellier 1997;
Mellier 1998; Jain, Seljak \& White 1999). 

%In summary, it appears that in the regime
%where the cosmic errors are small enough that the analysis
%is likely to have a significant power to constrain the
%underlying physics, 
%the theory of the cosmic errors and cross correlations, outlined
%in SC and SCB and thoroughly tested in paper I, is
%sufficient for maximum likelihood analysis.

\section*{Acknowledgments}
 We thank F. Bernardeau, P. Fosalba, C. Frenk, R. Scoccimarro,
 A. Szalay and S. White for useful discussions.
 It is a pleasure to acknowledge support for visits by IS and SC to the
 MPA, Garching and by SC to the dept Physics, Durham,
 during which part of this work was completed.           
 IS and AJ were supported by the PPARC rolling
 grant for
 Extragalactic Astronomy and Cosmology at Durham.

 The Hubble volume simulation data was made available by the Virgo
 Supercomputing Consortium
 (http://star-www.dur.ac.uk/$^{\sim}$frazerp/virgo/virgo.html).  The
 simulation
 was performed on the T3E at the Computing Centre of the Max-Planck
 Society in Garching. We would like to give our thanks to the many
 staff at the Rechenzentrum who have helped us to bring this project to
 fruition.                                                               

%
%
%========================

%

\bsp

\label{lastpage}

%
%===============
%===============
%===============
\end{document}